\documentstyle[twoside]{article}

\catcode`\@=11
\long\def\@makefntext#1{
\protect\noindent \hbox to 3.2pt {\hskip-.9pt  
$^{{\eightrm\@thefnmark}}$\hfil}#1\hfill}		

\def\@makefnmark{\hbox to 0pt{$^{\@thefnmark}$\hss}}	
	
\def\ps@myheadings{\let\@mkboth\@gobbletwo
\def\@oddhead{\hbox{}
\rightmark\hfil\eightrm\thepage}   
\def\@oddfoot{}\def\@evenhead{\eightrm\thepage\hfil
\leftmark\hbox{}}\def\@evenfoot{}
\def\sectionmark##1{}\def\subsectionmark##1{}}

\oddsidemargin=\evensidemargin
\newcounter{sectionc}\newcounter{subsectionc}\newcounter{subsubsectionc}
\renewcommand{\section}[1] {\vspace{12pt}\addtocounter{sectionc}{1} 
\setcounter{subsectionc}{0}\setcounter{subsubsectionc}{0}\noindent 
	{\tenbf\thesectionc. #1}\par\vspace{5pt}}
\renewcommand{\subsection}[1] {\vspace{12pt}\addtocounter{subsectionc}{1} 
	\setcounter{subsubsectionc}{0}\noindent 
	{\bf\thesectionc.\thesubsectionc. {\kern1pt \bfit #1}}\par\vspace{5pt}}
\renewcommand{\subsubsection}[1] {\vspace{12pt}\addtocounter{subsubsectionc}{1}
	\noindent{\tenrm\thesectionc.\thesubsectionc.\thesubsubsectionc.
	{\kern1pt \tenit #1}}\par\vspace{5pt}}
\newcommand{\nonumsection}[1] {\vspace{12pt}\noindent{\tenbf #1}
	\par\vspace{5pt}}

\newcommand{\textlineskip}{\baselineskip=13pt}
\newcommand{\smalllineskip}{\baselineskip=10pt}

\def\eightcirc{
\begin{picture}(0,0)
\put(4.4,1.8){\circle{6.5}}
\end{picture}}
\def\eightcopyright{\eightcirc\kern2.7pt\hbox{\eightrm c}} 

\newcommand{\copyrightheading}[1]
	{\vspace*{-2.5cm}\smalllineskip{\flushleft
        {\footnotesize Los Alamos electronic archives: quant-ph/9504017 #1}\\
        {\footnotesize $\eightcopyright$\, H.C. Rosu (1995) Phys. Lett. A 208
        (1995) 33-39
        }\\
	 }}

\newcommand{\publisher}[2]{{\begin{center}\footnotesize\smalllineskip 
	Received #1\\
	Revised #2
	\end{center}
	}}

\def\abstracts#1#2#3{{
	\centering{\begin{minipage}{4.5in}\baselineskip=10pt\footnotesize
	\parindent=0pt #1\par 
	\parindent=15pt #2\par
	\parindent=15pt #3
	\end{minipage}}\par}} 

\newcommand{\bibit}{\nineit}

\renewenvironment{thebibliography}[1]
	{\frenchspacing
	 \ninerm\baselineskip=11pt
	 \begin{list}{\arabic{enumi}.}
        {\usecounter{enumi}\setlength{\parsep}{0pt}     
	 \setlength{\leftmargin 12.7pt}{\rightmargin 0pt} 
         \setlength{\itemsep}{0pt} \settowidth
	{\labelwidth}{#1.}\sloppy}}{\end{list}}

\newcounter{itemlistc}
\newcounter{romanlistc}
\newcounter{alphlistc}
\newcounter{arabiclistc}

\def\@citex[#1]#2{\if@filesw\immediate\write\@auxout
	{\string\citation{#2}}\fi
\def\@citea{}\@cite{\@for\@citeb:=#2\do
	{\@citea\def\@citea{,}\@ifundefined
	{b@\@citeb}{{\bf ?}\@warning
	{Citation `\@citeb' on page \thepage \space undefined}}
	{\csname b@\@citeb\endcsname}}}{#1}}

\newif\if@cghi
\def\cite{\@cghitrue\@ifnextchar [{\@tempswatrue
	\@citex}{\@tempswafalse\@citex[]}}
\def\citelow{\@cghifalse\@ifnextchar [{\@tempswatrue
	\@citex}{\@tempswafalse\@citex[]}}
\def\@cite#1#2{{$\null^{#1}$\if@tempswa\typeout
	{IJCGA warning: optional citation argument 
	ignored: `#2'} \fi}}

\def\@refcitex[#1]#2{\if@filesw\immediate\write\@auxout
	{\string\citation{#2}}\fi
\def\@citea{}\@refcite{\@for\@citeb:=#2\do
	{\@citea\def\@citea{, }\@ifundefined
	{b@\@citeb}{{\bf ?}\@warning
	{Citation `\@citeb' on page \thepage \space undefined}}
	\hbox{\csname b@\@citeb\endcsname}}}{#1}}

\def\@refcite#1#2{{#1\if@tempswa\typeout
        {IJCGA warning: optional citation argument
	ignored: `#2'} \fi}}

\def\refcite{\@ifnextchar[{\@tempswatrue
	\@refcitex}{\@tempswafalse\@refcitex[]}}

\def\pmb#1{\setbox0=\hbox{#1}
	\kern-.025em\copy0\kern-\wd0
	\kern.05em\copy0\kern-\wd0
	\kern-.025em\raise.0433em\box0}

\def\fnt#1#2{\footnotetext{\kern-.3em
	{$^{\mbox{\scriptsize #1}}$}{#2}}}

\def\runninghead#1#2{\pagestyle{myheadings}
\markboth{{\protect\footnotesize\it{\quad #1}}\hfill}
{\hfill{\protect\footnotesize\it{#2\quad}}}}
\font\tenrm=cmr10
\font\tenit=cmti10 
\font\tenbf=cmbx10
\font\bfit=cmbxti10 at 10pt
\font\ninerm=cmr9
\font\nineit=cmti9

\font\eightrm=cmr8

\def\qed{\hbox{${\vcenter{\vbox{			
   \hrule height 0.4pt\hbox{\vrule width 0.4pt height 6pt
   \kern5pt\vrule width 0.4pt}\hrule height 0.4pt}}}$}}

\begin{document}

\runninghead{H.C. Rosu
$\ldots$} {H.C. Rosu
$\ldots$}


\normalsize\textlineskip
\thispagestyle{empty}
\setcounter{page}{1}

\copyrightheading{}			

\vspace*{0.88truein}

\centerline{\bf
Supersymmetry of Demkov-Ostrovsky effective potentials in the $R_{0}=0$
sector}
\vspace*{0.035truein}
\vspace*{0.37truein}
\centerline{\footnotesize H.C. Rosu (ifug, igss),
M.A. Reyes (cinvestav, ifug), K.B. Wolf (iimas-unam),
O. Obreg\'on (ifug, uam)}
\vspace*{0.015truein}
\baselineskip=10pt
\vspace*{10pt}
\vspace*{0.225truein}
\publisher{(April 20, 1995)}{(September 26, 1995)}

\vspace*{0.21truein}
\abstracts{
We present a supersymmetric analysis of the wave problem
with a Demkov-Ostrovsky spherically symmetric class of focusing potentials
at zero energy.
Following a suggestion of L\'evai, we work in the so-called $R_{0}=0$ sector
in order to obtain
the superpartner (fermionic) potentials within
Witten's supersymmetric procedure.
General solutions of the superpotential for the known physical cases are
given explicitly.
}{}{}


\textlineskip                  
\vspace*{12pt}                 

\vspace*{1pt}\textlineskip	
\vspace*{-0.5pt}
\noindent




\noindent







We consider the wave/quantum problem at fixed null energy, (or
zero-binding energy),
$[-\frac{\hbar ^2}{2m}
\nabla ^2_r + U_{\kappa}(r)]\psi ({\bf r})=0$, with Demkov-Ostrovsky (DO)
focusing potentials $^{1,2}$  
$$U_{\kappa}(r)=
-\frac{w{\cal E}_0}{(r/R)^2[(r/R)^{-\kappa}+(r/R)^{\kappa}]^2}, \eqno(1)$$
where $w>0$, $R>0$, $\kappa>0$ are constant parameters, and we introduced an
energy scale ${\cal E}_0=\hbar ^2/2mR^2$ for the potential part,
following Daboul and Nieto \cite{dn}.
Of the $U_{\kappa}$ class of potentials only the two cases to follow have
been studied in the literature. For $\kappa =1$, one
gets the wave problem for the Maxwell fish-eye (MF) lens, \cite{do}
whereas the $\kappa =1/2$ case
has been used for the atomic Aufbau (AA) chart since it fulfills the Madelung
rules of atomic energy ordering \cite{kb}. Demkov and Ostrovsky have shown
that for the cases $\kappa =k_1/k_2$, with $k_1$ and $k_2$ integers,
i) the classical
trajectories of a zero-energy (i.e., zero velocity at infinity) particle
close after $k_2$ revolutions around the force centre, and ii) all the
trajectories passing through a given point come to a focus after $k_2/2$
revolutions.

The above wave equation can be written
in the scaled variable $\rho =r/R$, (hereafter the energy scale is to be
understood), as follows

$$\Big[-\frac{\partial ^2}{\partial \rho ^2}-\frac{2}{\rho}
\frac{\partial}{\partial \rho}+ \frac{l(l+1)}{\rho ^2}-
\frac{w}{\rho ^{2(1-\kappa)}
(1+\rho ^{2\kappa})^2}\Big]\psi ({\bf \rho})=0.   \eqno(2)$$
It is quite straightforward to solve the
Sturm-Liouville problem, Eq.~(2). Moreover, it can be turned into an
eigenvalue
problem of the DO coupling constant, $w$, and also written as a Laplace
equation
on the four-dimensional sphere \cite{do}. The known results are the
following.
For $w$ taking the quantized values,
$w_{N,\kappa}=(2\kappa)^2[N+(2\kappa)^{-1}-1][N+(2\kappa)^{-1}]$,
the regular, normalizable solutions read
$$\psi _{Nlm}({\bf \rho})=R_
{Nl}(\rho)Y_{lm}(\theta, \phi),   \eqno(3a)$$
$$R_{Nl}(\rho)=\frac{{\cal N}_{Nl}}
{\rho ^{-l}(1+\rho ^{2\kappa})^{(2l+1)/2\kappa}}
C^{(2l+1)/2\kappa +1/2}_{N-1-l/\kappa}(\xi),   \eqno(3b)$$
where $\xi=\frac{1-\rho ^{2\kappa}}{1+\rho ^{2\kappa}}$,
$N=n +(\kappa ^{-1}-1)l$, $n=n_r +l+1$, $n_r =0,1,2,...$, are the
`total', `principal' and `radial' quantum numbers, $l$ and $m$ are the
spherical harmonic numbers, $C_{p}^{q}(\xi)$ are the
Gegenbauer polynomials,
i.e., the solutions of the corresponding ultraspherical equation
(see Eqs.~(7)
and (8) below), and ${\cal N}_{Nl}$ are the normalization constants.
What one gets when the $w$ parameter is made bigger and bigger is an increase
of the degeneracy of the normalizable state at zero-energy, but only for the
quantized values $w_{N,\kappa}$.
The degree of degeneracy of such a group of states is $N^2$,
($N$ = 1,2,3...), similar to the electron energy levels in a Coulomb field.

We pass now to
the functions $u_{Nl}=\rho R_{Nl}$ fulfilling the one-dimensional (half-line)
radial equation

$$H^-u\equiv
\Big[-\frac{\partial ^2}{\partial \rho ^2} + U_{eff}^{-}(\rho)\Big]u=0,
\eqno(4)$$
with the effective potential

$$U_{eff}^{-}= \frac{l(l+1)}{\rho ^2} -
\frac{(2\kappa)^{2}[N+(2\kappa)^{-1}-1][N+(2\kappa)^{-1}]}
{\rho ^{2(1-\kappa)}(1+\rho ^{2\kappa})^2},    \eqno(5)$$
where we have already included supersymmetric superscripts.
The functions $u_{Nl}$ are of the type
$f(\rho)C_{N-1-l/\kappa}^{(2l+1)/2\kappa +1/2}(\xi (\rho))$, where $f(\rho)$
reads
$$f(\rho)=\frac{\rho ^{l+1}}{(1+\rho ^{2\kappa})^{(2l+1)/2\kappa}},
\eqno(6)$$
and
the Gegenbauer polynomials, $C_{p}^{q}$, of degree $p=n_r$ and parameter $q$
as
given above, are the solutions of a
second-order differential (ultraspherical) equation of the type

$$P(\xi)\frac{d^2 C}{d\xi ^2}+Q(\xi)\frac{dC}{d\xi} +R_{p}(\xi)C(\xi)= 0,
\eqno(7)$$
with
$$P(\xi)=1,   \eqno(8.1)$$
$$Q(\xi)=\frac{2l+2\kappa +1}{\kappa}\frac{\xi}{\xi ^2 -1},  \eqno(8.2)$$
and
$$R_{p}(\xi)=-\frac{p(2q+p)}{\xi ^2 -1}.  \eqno(8.3)$$
In Eq.~(8.3), we emphasized the indexing of the $R$ functions according to
the various sectors $p=n_r$ (0,1,2,...), which is a well-known characteristic
of orthogonal polynomials \cite{lev}.

We have now all the requisites for a Natanzon-type approach that we outline
here following a recent discussion of L\'evai \cite{lev}.
The method has been first used by
Bhattacharjee and Sudarshan \cite{bs} and later by other authors, among whom
Natanzon \cite{nat} is the best known due to his systematic treatment of
hypergeometric cases.
The scheme deals with the fact that the solutions $\psi$ of any
one-dimensional Schr\"odinger equation can be written as $\psi(x)=
f(x)F(z(x))$,
where $f(x)$ is a function to be determined and directly related to the
superpotential, while
$F(z)$ is a special function which satisfies a second-order differential
equation of the form
$$\frac{d^2 F}{dz^2}+Q(z)\frac{dF}{dz} + R(z)F(z)=0.  \eqno(9)$$
In our case, $Q(z)$ and $R(z)$ corresponding to the Gegenbauer polynomials
have been written above, while $x=\rho$ and $z(x)=\xi (\rho)$.
Then, the following equations can be readily obtained
$$\frac{\xi ^{''}}{(\xi ^{'})^2} + \frac{2f^{'}}{\xi ^{'}f}=Q(\xi(\rho)),
\eqno(10)$$
and
$$\frac{f^{''}}{(\xi ^{'})^2 f}- \frac{U_{eff}}{(\xi ^{'})^2}= R(\xi (\rho)),
\eqno(11)$$
where $U_{eff}$ is given by Eq.~(5).
From Eq.~(10) the function $f(\rho)$ can be written as follows

$$f(\rho)\sim (\xi ^{'}(\rho))^{-1/2}\exp\Big[\frac{1}{2}
\int ^{\xi(\rho)}Q(\xi(\rho))d\xi\Big].  \eqno(12)$$

As suggested by L\'evai, \cite{lev} one can define a ground state
by the $R_{0}(\xi)=0$ sector, within which the Gegenbauer
polynomials are $C_{0}^{q}=1$ for any parameter $q$.
In this simple case, from Eq.~(11) one gets

$$U_{eff}=+W^2(\rho) -\frac{d W}{d\rho},  \eqno(13)$$
with $W(\rho)=-\frac{d}{d\rho}\ln f(\rho)$. Eq.~(13)
is just the initial Riccati equation for the DO cases (and for any radial
oscillator) in
Witten's supersymmetric quantum mechanics \cite{wit}. Hereafter Eq.~(13) will
be called DORE.
Since we actually know from Eq.~(6) the function $f(\rho)$ in the DO cases
(one can check that Eq.~(12) leads to the same function), a
short calculation gives the DO superpotential
$$W_{\kappa}(\rho)=\frac{l}{\rho}-\frac{2l+1}{\rho(1+\rho ^{2\kappa})}.
\eqno(14)$$
The effective DO superpartners in the $R_{0}=0$ sector can be written
as follows
$$U_{eff}^{-}=-\frac{dW_{\kappa}(\rho)}{d\rho}+W_{\kappa}^{2}(\rho),
\eqno(15a)$$
and
$$U_{eff}^{+}=+\frac{dW_{\kappa}(\rho)}{d\rho}+W_{\kappa}^{2}(\rho).
\eqno(15b)$$
Thus,
$$U_{eff}^{-}=\frac{l(l+1)}{\rho ^2}-
\frac{(2l+1)(2l+2\kappa +1)}{\rho ^{2(1-\kappa)} (1+\rho ^{2\kappa})^2},
\eqno(16a)$$
and
$$U_{eff}^{+}=\frac{l(l-1)}{\rho ^2}-
\frac{(2l+1)(2l-2\kappa -1)}
{\rho ^{2(1-\kappa)} (1+\rho ^{2\kappa})^2}
+\frac{2(2l+1)}{\rho ^{2}(1+\rho ^{2\kappa})^2}.
\eqno(16b)$$
The factorizing operators read
$$A=\frac{d}{d\rho} + W_{\kappa},   \eqno(17a)$$
and
$$A^{+}=-\frac{d}{d\rho}+ W_{\kappa}.  \eqno(17b)$$

We have plotted $U_{eff}^{-}$ and $U_{eff}^{+}$ for some values of the
parameters in Figs.~1 and 2.
From the plot of the DO fermionic potentials one can notice their
repulsive nature. Consequently, the fermionic equation
should be written in the continuum
$$H^{+}u_{1}\equiv AA^{+}u_{1}
\equiv(-\frac{d^2}{d\rho ^{2}}+U_{eff}^{+})u_{1}=
k^2 u_{1},\;\; k \in (0,\infty).   \eqno(18)$$
It will be investigated elsewhere. Here we remark that in order
to get the supersymmetric increment in the effective potential
we used only the particular solution of the Riccati equation coming
out from Eq.~(11). On the other hand,
it is well-known that the connection with the Gel'fand-Levitan
inverse scattering method requires the general solution of the Riccati
equation \cite{n}. We construct it in the usual two steps as follows.
Firstly,
consider ${\cal W}={\cal V} ^{-1} + W(\rho)$ as another solution.
Then by substituting in the DORE one gets
$$\frac{d{\cal V}}{d\rho}+2W(\rho) {\cal V}=-1~.
\eqno(19) $$
This equation can be written as follows
$$ \frac{d}{d\rho}\Bigg[
{\cal V}\exp\left( 2\int W(\rho) d\rho\right)\Bigg]
=-\exp \left(2\int W(\rho) d\rho\right)~,
\eqno(20)$$
with the solution
$${\cal V} =-\exp\left(
-2\int W(\rho) d\rho\right)\cdot \int \exp\left(2\int W(\rho) d\rho\right)
d\rho ~.
\eqno(21)$$
Since we know that $W(\rho)=-\frac{d}{d\rho}\ln f(\rho)$ we
get $\int W(\rho) d\rho =-\ln f(\rho)$. Thus
$${\cal V}=-f^2(\rho)\int f^{-2}(\rho)d\rho ~.
\eqno(22)$$
The object of interest is now the integral of the inverse square of $f$,
which reads explicitly
$$\int \frac{(\rho ^{-\kappa}+\rho ^{\kappa})^{\frac{(2l+1)}{\kappa}}}
{\rho}d\rho~ .
\eqno(23)$$
Let $\rho ^{\kappa}=\tan (\frac{\alpha}{2})$. Then the integral turns into
the form
$$\frac{2^{\frac{2l+1}{\kappa}}}{\kappa}
\int \frac{d\alpha}{(\sin \alpha)^{(2l+\kappa +1)/\kappa}}~,
\eqno(24)$$
and for the cases of physical interest, MF and AA, the formulas
2.515.1 and 2.515.2, respectively, in Gradshteyn and Ryzhik \cite{gr}
should be used to express it as a series.

Thus, in the MF case, ($\kappa$ =1), the integral Eq.~(24)
can be worked out into the series
$$S_1=-\frac{2^{2l+1}}{2l+1}\cos \alpha \left \{(\csc \alpha)^{2l+1}+
\sum _{m=1}^l \frac{2^m[l(l-1)...(l+1-m)]}{[(2l-1)(2l-3)...(2l+1-2m)]}
(\csc \alpha)^{2l+1-2m}\right\}~,
\eqno(25)$$
while for the AA case, ($\kappa$ =1/2), the integral Eq.~(24) reads
$$S_{\frac{1}{2}}=
-\frac{2^{4l+3}}{2l+1}\cos \alpha(\csc ^{2} \alpha)^{(2l+1)}
\Big[1 +
\sum_{m=1}^{2l}\frac{[(4l+1)(4l-1)...(4l-2m+3)]}
{(2\csc^{2} \alpha)^{m} [(2l)(2l-1)...(2l-m+1)]}
\Big]
+\frac{4[(4l+1)!!]}{4^{-l}[(2l+1)!]}\ln \tan (\frac{\alpha}{2})~,
\eqno(26)$$
where $\alpha =2\arctan \rho$ in the first formula and $\alpha=2\arctan
\sqrt{\rho}$ in the latter one.

In the MF case, the radial factor $f^2(\rho)$ can be written
trigonometrically as
$2^{-2l}(\sin\alpha)^{2l}\sin ^2(\frac{\alpha}{2})$ implying
$${\cal V} _{1}=\frac{2\cos \alpha}{2l+1}\tan (\frac{\alpha}{2})
\left\{1+
\sum _{m=1}^l \frac{(2\sin^2 \alpha)^m
[l(l-1)...(l-m)]}{[(2l-1)(2l-3)...(2l+1-2m)]}\right\}~.
\eqno(27)$$

In the AA case, the square radial factor is
$2^{-4l}(\sin \alpha)^{4l}\sin ^{4} (\frac{\alpha}{2})$ and
one can work out easily a formula for ${\cal V} _{\frac{1}{2}}$
$$
{\cal V} _{\frac{1}{2}}=  \frac{2\cos \alpha}{2l+1}
\tan ^2 (\frac{\alpha}{2})
\Big[1+
\sum_{m=1}^{2l}\frac{(\frac{\sin^2 \alpha}{2})^m
[(4l+1)(4l-1)...(4l-2m+3)]}{(2l)(2l-1)...(2l-m+1)}\Big]
+ \frac{4[(4l+1)!!][\frac{\sin ^{2}(\alpha)}{2}]^{2l}}
{(2l+1)![\csc (\frac{\alpha}{2})]^{4}}
\ln \tan (\frac{\alpha}{2})~.
\eqno(28)$$
We have in this way all the ingredients for the second step, which is to
write down the general solution of DORE,
containing a constant $\lambda$-parameter in Eq.~(22),
see Refs. $^{8,10}$.
The general solution means Eq.~(22) modified as follows
$$ {\cal V} _{\lambda,\kappa}=-f^2_{\kappa}(\rho)
\left(\lambda +\int _{\rho}^{\infty}f^{-2}_{\kappa}
(\rho ^{'})d\rho ^{'}\right)~.
\eqno(29) $$
In this case, the integral equation (24) will have the lower
limit $\alpha$ and the upper
one $\pi$. Thus, ${\cal V} _{\lambda, \kappa}$ can be calculated from our
formulas as
${\cal V} _{\lambda, \kappa}={\cal V} _{\kappa}-
\lambda f^{2}_{\kappa}(\rho)$.


\newpage

\nonumsection{Acknowledgements}
\noindent
This work was partially supported by the CONACyT Projects 4862-E9406,
4868-E9406 and by the Project DGAPA IN 1042 93 at the Universidad
Nacional Autonoma de M\'exico.
M.R. was supported by a CONACyT Graduate Fellowship.

\vskip 0.5cm

{\bf Figure Captions}
\bigskip

{\bf Fig. 1}.
Effective DO superpartners in the $R_{0}=0$ sector,
$U_{eff}^{\mp}(\rho)$, (a) and (b),
respectively, in units of ${\cal E}_{0}$, for  $l=2$, and $\kappa$ =
1/2, 1, and 3/2.

{\bf Fig. 2}.
Effective superpartners of the $R_{0}=0$ sector in ${\cal E} _0$ units:
(a), $U^{-}_{eff}$ for $l$= 1, 5, 10,
and (b), $U^{+}_{eff}$ for
$l$=6, 7, 8, in the case of Maxwell fish-eye, $\kappa =1$. We have
plotted $U^{+}_{eff}$ in the region of the critical (inflexion) angular
number, $l_{cr}$, that
we have found numerically to be $l_{cr}$=6.876 for $\rho _{cr}$=1.599. The
critical $l$ is the entry point toward a pocket (trapping) region of
$U^{+}_{eff}$ for $l>l_{cr}$.

\vskip 0.5cm


{\bf Note added on January 21st, 1996}

\bigskip

The mathematics in the last two pages of the paper (pp. 38, 39), though not
wrong, might be considered as misleading with respect to the literature,
since we obtained the general solution of the initial
`bosonic' Riccati equation. In this way, one can introduce the one-parameter
family of fermionic potentials with the same bosonic superpartner. This is
what we have done. However, for the usual connection with the
Gel'fand-Levitan method, one
should obtain the general solution of the `fermionic' Riccati equation,
and thus, the one-parameter family of bosonic potentials with the same
fermionic superpartner. The task can be readily done on the base of the
results at pp. 38, 39.




\nonumsection{References}




\begin{thebibliography} {000}

\bibitem{do}
Yu.N. Demkov and V.N. Ostrovsky, JETP 33 (1971) 1083.

\bibitem{kb}
Y. Kitagawara and A.O. Barut, J. Phys. B 16 (1983) 3305; 17 (1984) 4251;
Yu.N. Demkov and V.N. Ostrovsky, Zh. Eksp. Teor. Fiz. 62 (1971) 125 [Sov.
Phys. JETP 35 (1972) 66]; 63 (1972) 2376E; V.N. Ostrovsky, J. Phys. B 14
(1981) 4425; V.N. Ostrovsky, in {\bibit Latin-American School of Physics
XXX ELAF}, eds. O. Casta\~nos {\em et al}, AIP Conf. Proc. No. 365
(AIP, New York, 1996) pp. 191-216; V.N. Ostrovsky, Phys. Rev. A 56 (1997)
626.

\bibitem{dn}
J. Daboul and M.M. Nieto, Phys. Lett. A 190 (1994) 357.

\bibitem{lev}
G. L\'evai, J. Phys. A 22 (1989) 689.

\bibitem{bs}
A. Bhattacharjee and E.C.G. Sudarshan, Nuovo Cimento 25 (1962) 864.

\bibitem{nat}
G.A. Natanzon, Teor. Mat. Fiz. 38 (1979) 146.

\bibitem{wit}
E. Witten, Nucl. Phys. B 185 (1981) 513.

\bibitem{n}

M.M. Nieto, Phys. Lett. B 145 (1984) 208;
B. Mielnik, J. Math. Phys. 25 (1984) 3387.


\bibitem{gr}
I.S. Gradshteyn and I.M. Ryzhik, {\em Table of Integrals, Series, and
Products}, 4th edition, (Academic Press, 1980) pp. 134.

\bibitem{suk}
C.V. Sukumar, J. Phys. A 18 (1985) 2917. See the Appendix therein.






\end{thebibliography}
\end{document}